\documentclass[aps,prb,twocolumn,showpacs,superscriptaddress]{revtex4-1}

\let\oldhat\hat
\renewcommand{\hat}[1]{\oldhat{\mathbf{#1}}}

\usepackage{amsmath,amsfonts}
\usepackage{graphicx}

\usepackage{dcolumn}
\usepackage{longtable}
\usepackage{color}
\usepackage{setspace}

\begin{document}
\title{Origins of the structural phase transitions in MoTe$_2$ and WTe$_2$}

\author{Hyun-Jung Kim}
\affiliation{Korea Institute for Advanced Study, Hoegiro 85, Seoul 02455, Korea}

\author{Seoung-Hun Kang }
\affiliation{Korea Institute for Advanced Study, Hoegiro 85, Seoul 02455, Korea}

\author{Ikutaro Hamada}
\altaffiliation[Present Address: ]{Department of Precision Science and Technology,
Graduate School of Engineering, Osaka University, Osaka 565-0871, Japan}
\affiliation{Center for Green Research on Energy and Environmental Materials, 
National Institute for Materials Science, Tsukuba 305-0044, Japan}

\author{Young-Woo Son}
\email{E-mail: hand@kias.re.kr}
\affiliation{Korea Institute for Advanced Study, Hoegiro 85, Seoul 02455, Korea}

\date{\today}

\begin{abstract}
Layered transition metal dichalcogenides MoTe$_2$ and WTe$_2$ share almost similar lattice constants as well as topological electronic properties except their structural phase transitions. While the former shows a first-order phase transition between monoclinic and orthorhombic structures, the latter does not. Using a recently proposed van der Waals density functional method, we investigate structural stability of the two materials and uncover that the disparate phase transitions originate from delicate differences between their interlayer bonding states near the Fermi energy. By exploiting the relation between the structural phase transitions and the low energy electronic properties, we show that a charge doping can control the transition substantially, thereby suggesting a way to stabilize or to eliminate their topological electronic energy bands.
\end{abstract}
\pacs{73.22.-f, 71.15.Mb, 64.70.Nd}

\maketitle


Since the successful exfoliation of various two dimensional (2D) crystals in 2005~\cite{Novoselov2005}, the layered materials in a single layer as well as bulk forms have attracted serious attention owing to their versatile physical properties~\cite{Gupta2015,Geim2013}. Among them, the layered transition metal dichalcogenides (TMDs) show various interesting electronic properties such as type-II Weyl semimetallic (WSM) energy bands~\cite{Soluyanov2015}, gate dependent collective phenomena~\cite{Ye2012,Yu2015}, and quantum spin Hall (QSH) insulating state~\cite{Qian2015} to name a few.

Because of the layered structures of TMDs, several polymorphs can exist and show characteristic physical properties depending on their structures~\cite{Kolobov2016}. A typical TMD shows the trigonal prismatic (2$H$) or the octahedral (1$T$) structures~\cite{Mak2010,Wang2012,Mattheiss1973,Wilson1969}. For MoTe$_2$ and WTe$_2$, the 2$H$ structure ($\alpha$-phase, $P$6$_{3}$/$mmc$) is a stable semiconductor while the 1$T$ form is unstable~\cite{Qian2015,Keum2015}. The unstable 1$T$ structure turns into the distorted octahedral one (1$T'$)~\cite{Eda2012,Qian2015}. The stacked 1$T'$ single layer forms a three-dimensional bulk with the monoclinic structure ($\beta$-phase, $P$2$_1$/$m$) or the orthorhombic one ($\gamma$-phase, $P$$mn$2$_1$) (see Fig. 1)~\cite{Brown1966,Dawson1987,Mar1992}. Interestingly, the $\beta$ phase with a few layers is a potential candidate of QSH insulator~\cite{Qian2015} and the bulk $\gamma$ phase shows type-II Weyl semimetalic energy bands~\cite{Sun2015,Soluyanov2015,Chang2016}, respectively. Since the structural differences between $\beta$ and $\gamma$ phases are minute ($\sim$4$^\circ$ tilting of axis along out-of-plane direction in $\beta$ phase with respect to one in $\gamma$ phase), the sensitive change in their topological low energy electronic properties is remarkable and the transition between different structures can lead to alternation of topological properties of the system.

\begin{figure}[b]
\includegraphics[width=0.8\columnwidth]{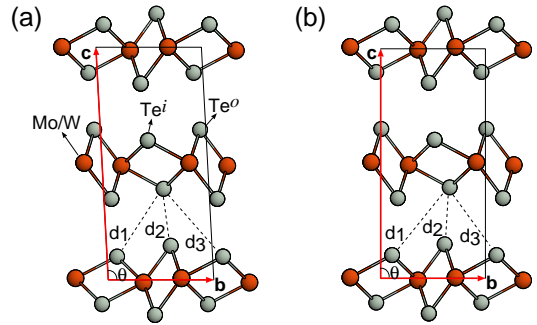}
\caption{(Color online) Schematic atomic structures of (a) the $\beta$ and (b) the $\gamma$ phase of MoTe$_2$ and WTe$_2$  projected on the $bc$ plane. $\bf b$ and $\bf c$ denote unit vectors of the primitive unit cell ($\bf a$ is perpendicular to the $bc$ plane). The solid line indicates the unit cell. The dark (red) and bright (gray) circles represent Mo (W) and Te atoms, respectively. Te atom being close to (away from) the transition metal plane is denoted by Te$^{i(o)}$, respectively. For the $\beta (\gamma)$ phase, $d_1$$<$$d_3$ ($d_1$$>$$d_3$). The angle between $\bf b$ and $\bf c$ is (a) $\theta$ $\simeq$ 94$^\circ$ and (b) $\theta$ = 90$^\circ$.}
\end{figure}

A phase transition between the $\beta$- and $\gamma$-phase in the layered TMDs has been known for a long time~\cite{Clarke1978,Dawson1987}. MoTe$_2$ shows a first-order transition from the $\beta$- to $\gamma$-structure at around 250 K~\cite{Clarke1978} when temperature decreases. WTe$_2$, however, does not show any transition and stays at the $\gamma$-phase~\cite{Kang2015,Pan2015}. Since the structural parameters of a single layer of 1$T'$-MoTe$_2$ and 1$T'$-WTe$_2$ are almost the same~\cite{Brown1966,Mar1992,Choe2016} and Mo and W belong to the same group in the periodic table, the different phase transition behaviors are intriguing and origins of the contrasting features are yet to be clarified.

To understand the phase transition, the proper treatment of long and short range interlayer interaction in TMDs is essential. Most of the theoretical studies, however, fail to reproduce the experimental crystal structures of the two phases of MoTe$_2$ and WTe$_2$ so do their topological electronic structures using crystal structures obtained from $ab$ $initio$ calculations~\cite{Lee2015,Zhao2015,Lv2015,Homes2015,Liu2016,Qi2016,Lu2016}. Instead, the atomic structures from experiment data are routinely used to understand and predict the low energy electronic properties~\cite{Soluyanov2015,Sun2015,Wang2016,Tamai2016,Deng2016,Huang2016,Bruno2016}. This is because the calculated lattice parameters, especially interlayer distance, by using the conventional first-principles calculations~\cite{Lee2015,Zhao2015,Chang2016,supp} [even with advanced empirical van der Waals (vdW) interaction correction schemes~\cite{Qi2016,supp,Lee2015}] hardly reproduce the observed distances. Since the interlayer interaction governs the phase transition as well as structural properties, a successful description of interlayer interactions is required to understand or predict electronic structures and topological properties. Motivated by the current situation of experiment and theoretical studies, we perform $ab$ $initio$ calculations using a new vdW density functional method for the interlayer interaction~\cite{Hamada2014} and analyze the existence and absence of the first-order structural phase transition related with various low energy topological electronic properties of MoTe$_2$ and WTe$_2$.

Here we first compute crystal structures of the both compounds based on an advanced self-consistent density functional method for the vdW interaction~\cite{Hamada2014} and obtain the best agreement with the available crystal structures in experiments. Then we show theoretically that MoTe$_2$ and WTe$_2$ have distinct structural phase transitions because their interlayer bondings differ depending on valence electron configurations of transition metals. A critical role of low energy electronic states for crystal symmetry is further demonstrated by showing that an external charge doping can alter the structural phase transition significantly. From this, our results in this Rapid Communication can provide a firm computational and theoretical basis for future development in discovering and engineering various topological electronic states in layered materials.

\begin{figure}[t]
\centering{ \includegraphics[width=1.0\columnwidth]{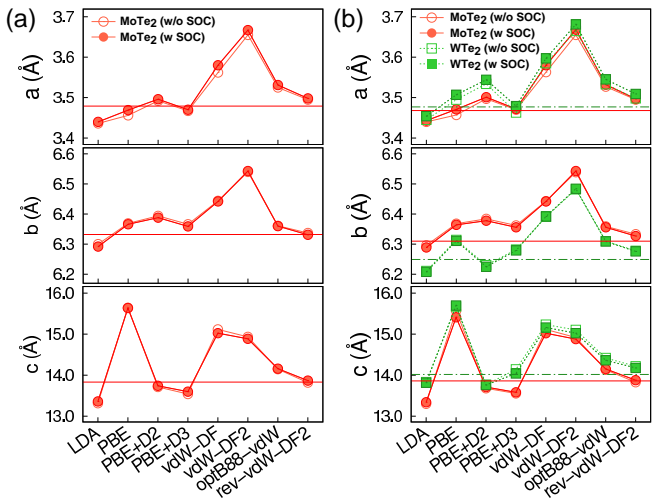} }
\caption{(Color online) Optimized lattice parameters $a$, $b$, and $c$ for (a) the $\beta$ and (b) the $\gamma$ structures, obtained using different exchange-correlation functionals. Experimental values for $\beta$-MoTe$_2$ and $\gamma$-MoTe$_2$ (Ref.~\cite{Tamai2016})  and those for $\gamma$-WTe$_2$(Ref.~\cite{Mar1992}) are shown by horizontal solid lines, and horizontal dotted lines, respectively.}
\end{figure}

Our {\it ab initio} calculation method employs the projector-augmented wave method~\cite{PAW} as implemented in the Vienna Ab-initio Simulation Package (${\rm VASP}$)~\cite{VASP1,VASP2}. We use the plane-wave cutoff of 450 eV and the 32$\times$16$\times$8 Monkhorst-Pack meshes for the Brillouin zone integration to achieve the convergence criterion of 0.1 meV in total energy difference ($\Delta E_{\gamma-\beta}$) between $\beta$ and $\gamma$ phase. The spin-orbit coupling (SOC) effect is included in all calculations and on-site Coulomb repulsion ($U$)~\cite{Dudarev1998} is considered for the specific cases. These parameters are fully tested to achieve a desired accuracy for the calculations, and the energy and force are converged with thresholds of 10$^{-6}$ eV and 5$\times$10$^{-3}$ eV/\AA, respectively. On top of the conventional calculation method, we use a vdW density functional (rev-vdW-DF2) method which is recently proposed by one of the authors~\cite{Hamada2014}, where the revised Becke exchange functional (B86R)~\cite{Becke1986} is adopted for exchange functional together with the second version of nonlocal vdW-DF (vdW-DF2)~\cite{Dion2004,*Dion2004e,Lee2010} as a nonlocal correlation. The rev-vdW-DF2 improves the description of the attractive vdW interaction resulting in the most accurate interlayer distances of layered materials over the various other vdW calculation methods~\cite{supp,Bjorkman2014,Peng2016}. The electron and hole dopings are simulated by adding and removing the electron and the background charge is added to keep the charge neutrality. To evaluate the vibrational energy and entropy, we use the harmonic approximation as implemented in PHONOPY package~\cite{phonopy} where the vibrational frequencies are obtained from the force constant matrix of the fully relaxed geometries using numerical derivatives of the rev-vdW-DF2 energies.

The atomic structures obtained from our calculation match the available experiment data very well. The calculated structural parameters of MoTe$_2$ in the $\beta$ phase (hereafter called $\beta$-MoTe$_2$) are summarized in Fig. 2 (a) and those for MoTe$_2$ and WTe$_2$ in the $\gamma$ phase ($\gamma$-MoTe$_2$ and $\gamma$-WTe$_2$) are summarized in Fig. 2 (b) (see also Tables SI and SII~\cite{supp}). The comparison between the optimized lattice parameters using the various vdW functionals and experiment data for the $\beta$ and $\gamma$ phases are also illustrated, respectively. We note that the inclusion of SOC improves the accuracy marginally (see Fig. 2, Tables SI and SII~\cite{supp}). Among the various vdW correction schemes, we found that the rev-vdW-DF2 outperforms several other functionals. The calculated equilibrium unit cell volume using our method yields 306.5 \AA$^3$ for the $\beta$-MoTe$_2$ and 307.0 and 312.1 \AA$^3$ for the $\gamma$-MoTe$_2$ and $\gamma$-WTe$_2$, respectively, in very good agreement with experimental value of 303.6, 305.9, and 306.6 \AA$^3$, respectively. These are only larger by 1.0, 0.4, and 1.8 ${\%}$ than those from experiment, respectively. From the fully optimized structures for both phases, we find that the shortest interlayer distance between Te atoms (denoted by $d_2$ in Fig. 1) changes negligibly between the two phases while other distances ($d_1$ and $d_3$) vary significantly (see Fig. 1 and Table SIII~\cite{supp}).

\begin{figure}[b]
\centering{ \includegraphics[width=1.0\columnwidth]{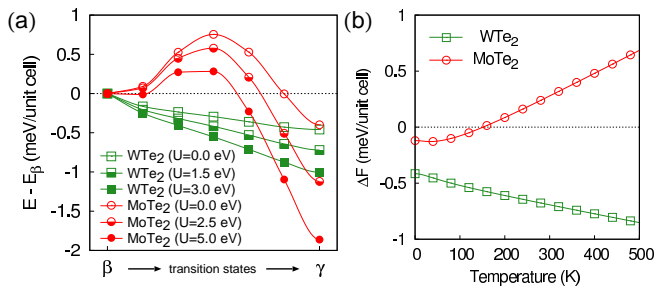} }
\caption{(Color online) (a) Energy profile calculated using rev-vdW-DF2 with and without SOC and $U$ along the transition path from $\beta$- to $\gamma$-phase of MoTe$_2$ and WTe$_2$ with respect to the total energy of the $\beta$-phase. (b) Calculated free energy difference $\Delta F=F_{\gamma}-F_{\beta}$  using rev-vdW-DF2 with SOC  and without $U$, where $F_{\gamma(\beta)}$ is a free energy of $\gamma(\beta)$ phase.}
\end{figure}

As the temperature increases, the stable $\gamma$-MoTe$_2$ at the low temperature undergoes a first-order phase transition to the $\beta$ phase~\cite{Clarke1978,Dawson1987,Mar1992} while WTe$_2$ stays in the $\gamma$ phase~\cite{Kang2015,Pan2015}. These observations are consistent with our total energy calculation including the vdW interaction. We found that the $\gamma$ phase is energetically more stable than the $\beta$ phase by $\Delta E_{\gamma-\beta}$ = 0.40 and 0.46 meV per unit cell for MoTe$_2$ and WTe$_2$, respectively, in good agreement with recent other studies~\cite{Qi2016,Lu2016}. For MoTe$_2$, the transition state  is unstable by 0.75 and 1.15 meV per unit cell than the $\beta$- and $\gamma$-phase, respectively, indicating $\beta$-MoTe$_2$ is metastable state, while WTe$_2$ shows no energy barrier, implying that $\beta$-WTe$_2$ does not exist [see Fig. 3(a)]. An atomic structure of the hypothetical $\beta$-WTe$_2$ is assumed to follow $\beta$-MoTe$_2$. We also calculated the free energy of each system without $U$ and found that the structural phase transition occurs at around 150 K for MoTe$_2$ and no transition for WTe$_2$, compatible with the experiment [Fig. 3(b)].

Recent studies~\cite{Keum2015,Zheng2016} show that the insulating behavior of a few layers of MoTe$_2$ and WTe$_2$ are not described well within the mean-field treatment of Coulomb interactions. This implies a critical effect of many-body interaction. Thus, we further add the local Coulomb repulsion of $U$ on top of our rev-vdW-DF2 method to reproduce the finite energy band gap obtained from previous hybrid density functional calculations~\cite{Keum2015,supp}. We set $U$ to be 5.0 and 3.0 eV for Mo 4$d$ and W 5$d$ orbitals, respectively~\cite{supp} and obtain further increasing $\Delta E_{\gamma-\beta}$ = 1.9 and 1.0 meV per unit cell for MoTe$_2$ and WTe$_2$, respectively. We note that inclusion of $U$ stabilizes the $\gamma$ phase of both materials while the transition energy barrier for MoTe$_2$ decreases with increasing $U$ [Fig. 3(a)].

\begin{figure}[t]
\centering{ \includegraphics[width=1.0\columnwidth]{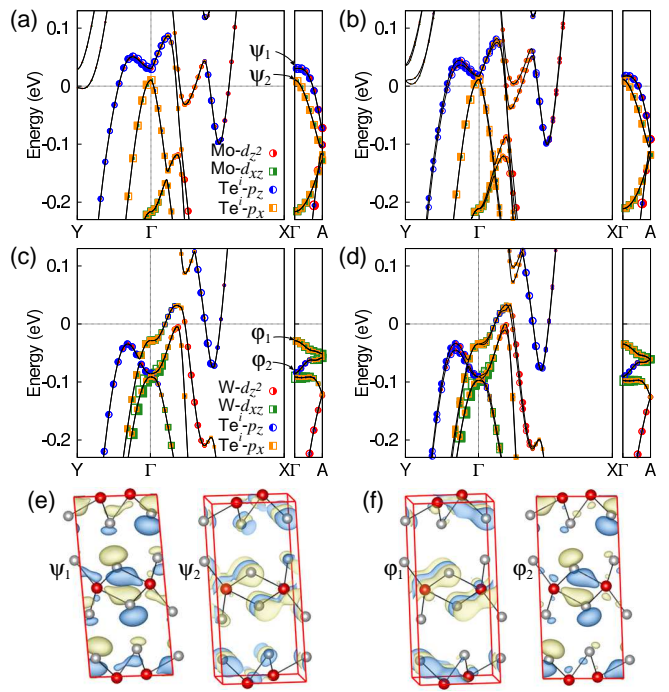} }
\caption{(Color online) Band structures of (a) $\beta$-MoTe$_2$, (b) $\gamma$-MoTe$_2$, (c) $\beta$-WTe$_2$, and (d) $\gamma$-WTe$_2$ using rev-vdW-DF2 method with SOC. The Fermi energy ($E_F$) is set to zero. The bands are plotted along $Y$(0,$\frac{1}{2}$,0)$\rightarrow$$\Gamma$(0,0,0)$\rightarrow$$X$($\frac{1}{2}$,0,0) and $\Gamma$(0,0,0)$\rightarrow$$A$(0,0,$\frac{1}{2}$). The bands projected onto the $d_{xz}$ and $d_{z^2}$ orbitals of Mo and $p_x$ and $p_z$ orbitals of Te are displayed with circles whose radii are proportional to the weights of each orbital. To visualize the bonding nature of valence bands, the wave functions at the $\Gamma$ point are drawn for (e) $\psi_1$ and $\psi_2$ of $\beta$-MoTe$_2$ and (f) $\varphi_1$ and $\varphi_2$ of $\beta$-WTe$_2$ where blue (green) color denotes plus (minus) sign.}
\end{figure}

In Fig. 4, we show the low energy electronic bands near the Fermi energy ($E_F$) for two different phases of MoTe$_2$ and WTe$_2$, respectively. We first find that the two compounds show the markedly different band dispersion along the $\Gamma$-$A$ direction. For MoTe$_2$, the topmost partially occupied valence band state [denoted by $\psi_1$ in Figs. 4(a) and 4(e)] is mainly an antibonding state along the $d_1$ direction (see Fig. 1) between the hybridized states of $p_z$ orbital of the lower Te atom (denoted by Te$^i$ in Fig. 1) and $d_{z^2}$ orbital of Mo. The next valence band state [$\psi_2$ in Fig. 4(a)] is mainly an antibonding state between the hybridized states of $p_x$ orbital of Te$^i$ and $d_{xz}$ orbital of Mo [Fig. 4(e)]. We also note that, in the first two valence bands, contribution of $p$ orbitals of Te$^o$ (see Fig. 1) is relatively smaller than those of Te$^i$. In contrast to the case of MoTe$_2$, the topmost valence state [$\varphi_1$ state in Figs. 4(c) and 4(f)] of WTe$_2$ is similar to the second valence state ($\psi_2$) of MoTe$_2$ and vice versa [Fig. 4(f)]. Because of the different atomic orbital configurations between Mo ([Kr]5s$^1$4d$^5$) and W atom ([Xe]6s$^2$4f$^{14}$5d$^4$), those two valence bands of WTe$_2$ are fully occupied along the $\Gamma$-$A$ and $\Gamma$-$Y$ direction [Figs. 4(b) and 4(d)] while those of MoTe$_2$ are partially occupied along all directions [Figs. 4(a) and 4(c)]. The estimated band width along $\Gamma$-$A$ for those two bands of MoTe$_2$ is four times larger than the width of WTe$_2$. These apparent differences between the two compounds are found to originate from the fact that WTe$_2$ has a quite smaller contribution of $p$ orbital of Te atoms to the first two valence states compared to that of MoTe$_2$ (Fig. S2~\cite{supp}]). 
We also calculated the whole band structures again using a semilocal correlational functional (Fig. S3~\cite{supp}) instead of the rev-vdW-DF2 while keeping the fully relaxed atomic structures to check the effect of vdW functional on the energy band structures. Changes in the band structures are found to be minimal agreeing with previous studies~\cite{Thonhauser2007, Hamada2011}.

Since the calculated total energy difference between the two phases is very small, we do not expect significant changes between energy bands of different phases. Indeed, as shown in Figs. 4 and S4~\cite{supp}, there are little modifications in the band structures between the two phases of MoTe$_2$ (WTe$_2$) except that all bands in the $\beta$ phase split into spin-polarized ones in the $\gamma$ phase due to its broken inversion symmetry. However, in MoTe$_2$, there is a small but important variation in the band structures with the transition: The partially occupied valence bands related with the interlayer antibonding states ($\psi_1$ and $\psi_2$) in the $\beta$-MoTe$_2$ move down in energy (are steadily occupied) along the transition pathway to the $\gamma$-MoTe$_2$ while the corresponding states in the $\beta$-WTe$_2$ does not (Figs. 4 and S4~\cite{supp}). The increase in the occupancies in the first two valence bands stabilize the antibonding states along the elongated distance of $d_1$~\cite{Kim2015}. This is made possible because there is a net charge transfer from the intralayer bonding states around the $Y$-point to the interlayer anti-bonding states near the $E_F$ as shown in Fig. S4~\cite{supp}. This costs energy and explains the metastability of the $\beta$-MoTe$_2$. Since those bands in WTe$_2$ are all occupied, there is no metastable phase for the WTe$_2$.

\begin{figure}[t]
\centering{ \includegraphics[width=1.0\columnwidth]{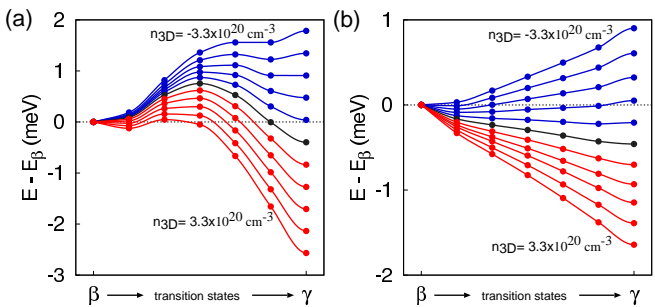} }
\caption{(Color online) Calculated energy profile (with SOC and without $U$) along the transition path from $\beta$- to $\gamma$-phase of (a) MoTe$_2$ and (b) WTe$_2$ as a function of doping ($n_{3D}$) ranging from $-$3.3$\times$10$^{20}$cm$^{-3}$ to +3.3$\times$10$^{20}$cm$^{-3}$. The energy profiles with electron (positive) doping, hole (negative) doping and neutral case are drawn by red, blue, and black lines, respectively. Doping density difference between the consecutive lines is 6.6$\times$10$^{19}$ cm$^{-3}$.} 
\end{figure}

Considering the crucial role of occupancy of the interlayer bonding states near the $E_F$, we expect that the external doping can control the structural phase transition. Indeed, we find that the hole (electron) doping can stabilize the $\beta$($\gamma$) phase of both compounds as shown in Fig. 5.  The amount of doping density that is necessary to invert the direction of phase transition is about 1.0$\times$10$^{20}$ cm$^{-3}$. We note that few recent experiments~\cite{Ye2012,Yu2015} can achieve such a level of doping for thin TMD flakes. It is anticipated that the $in$-$situ$ charge or hole injection can turn on and off QSH insulating phase and WSM states, respectively. We also note that only electron doping can push the $E_F$ to the Weyl points of WSM states because hole doping destroys the $\gamma$ phase.

Lastly, we comment on the existence of Weyl points calculated from our $ab$ $initio$ atomic structures of both compounds. For $\gamma$-MoTe$_2$ and $\gamma$-WTe$_2$, all the bands are split into spin polarized ones thanks to the broken inversion symmetry and SOC [Figs. 4(b) and 4(d)]. As already shown by other studies~\cite{Sun2015,Tamai2016}, we also find eight Weyl points of $\gamma$-MoTe$_2$ in the $k{_z}$ = 0 plane (see Fig. S5~\cite{supp}). Unlike the robust Weyl points in $\gamma$-MoTe$_2$, the slight overestimation of $a$ and $c$ axes (by 0.5 and 1.0$\%$) in our calculation for $\gamma$-WTe$_2$ [see Fig. 2(b)] merges the topological Weyl points with the opposite chiralities~\cite{Soluyanov2015}, highlighting their sensitivity on the detailed structure parameters. We can recover the eight Weyl points in the $k_{z}$ = 0 plane of $\gamma$-WTe$_2$ under biaxial strain ($\bf a$ and $\bf c$) of $-$1.5$\%$ (see Fig. S6~\cite{supp}). 

In conclusion, using an advanced $ab$ $initio$ calculation method for the vdW interaction, we computed accurate lattice structures of MoTe$_2$ and WTe$_2$ and uncovered origins of their disparate structural phase transition phenomena. We showed that the slight differences in low energy states related with the interlayer bondings are shown to be pivotal in determining the symmetry of bulk crystals. Since the structural transition intertwines their QSH phase and WSM states, our results shed light onto understanding delicate interplay between topological electronic properties and crystal structures. Furthermore, we find that the electron and hole doping alter the structural phase transitions, opening a way to control the topological electronic properties of layered TMDs using available experiment techniques.

We thank Dr. Jun-Ho Lee for fruitful discussions at an early stage of this work. Y.-W.S. was supported by the National Re- search Foundation of Korea funded by the Ministry of Science, ICT and Future Planning of Korean government (QMMRC, No, R11-2008-053-01002-0). The computing resources were supported by the Center for Advanced Computation (CAC) of KIAS.
 
\newpage
\twocolumngrid
%

\newpage

\onecolumngrid

\section{Supplemental Figures and Tables}

\renewcommand{\thefigure}{S\arabic{figure}}
\renewcommand{\thetable}{S\Roman{table}}

\setcounter{figure}{0}

\begin{figure}[h]
\centering{ \includegraphics[width=16.0cm]{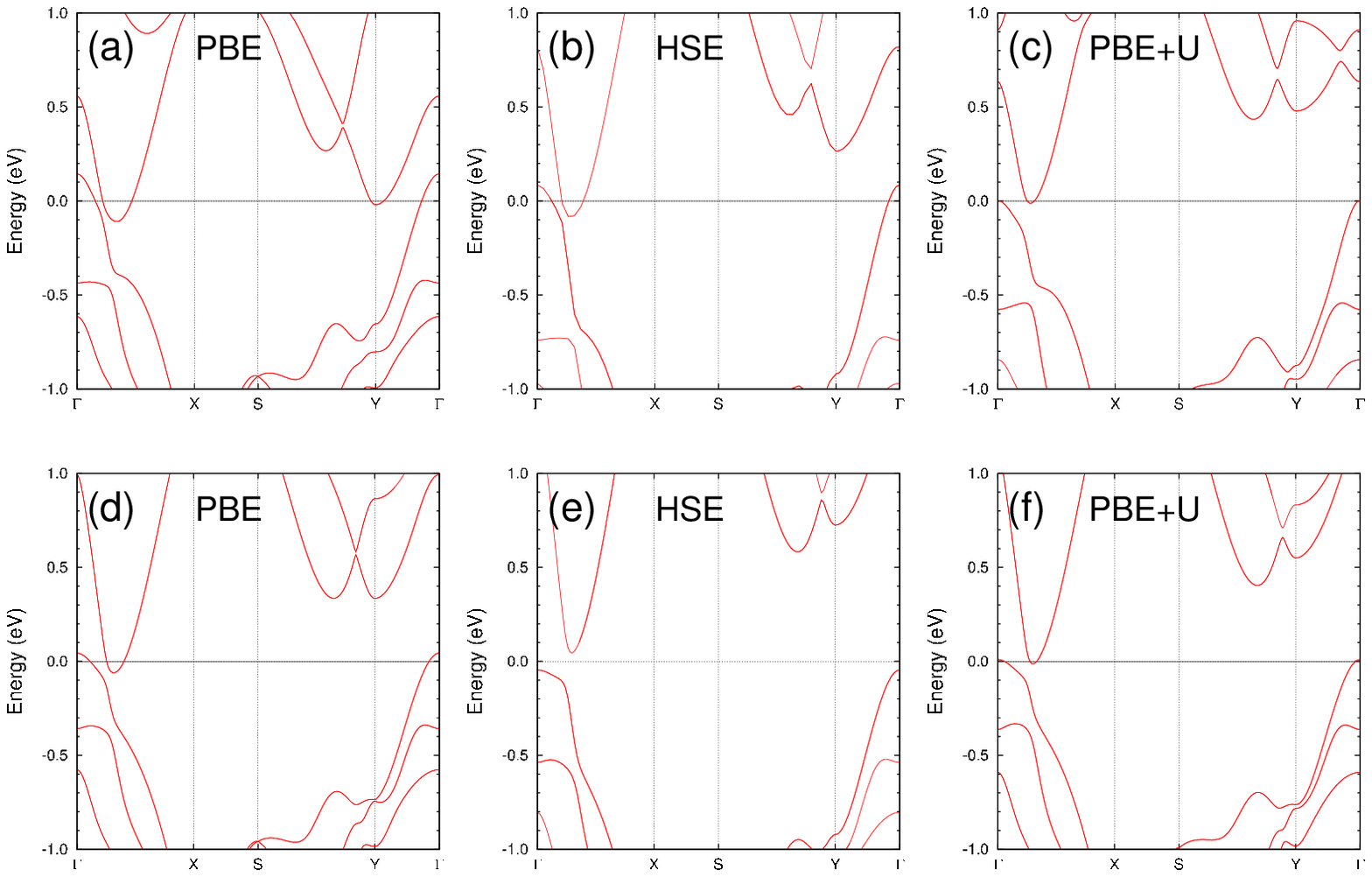} }
\caption{(Color online) Band structures of the monolayer (a)-(c) 1$T'$-MoTe$_2$ and (e)-(f) 1$T'$-WTe$_2$ obtained by PBE, HSE, and PBE+$U$ functionals including SOC effect. The atomic structures are fully relaxed with each functional. We set $U$ to be (c) 5.0 and (f) 3.0 eV for Mo 4$d$ and W 5$d$ orbitals, respectively. The Fermi energy ($E_F$) is set to zero. The band structures are plotted along the path $\Gamma$(0,0,0)$\rightarrow$$X$($\frac{1}{2}$,0,0)$\rightarrow$$S$($\frac{1}{2}$,$\frac{1}{2}$,0)$\rightarrow$$Y$(0,$\frac{1}{2}$,0)$\rightarrow$$\Gamma$(0,0,0).}
\end{figure}

\begin{figure}[h]
\centering{ \includegraphics[width=0.75\columnwidth]{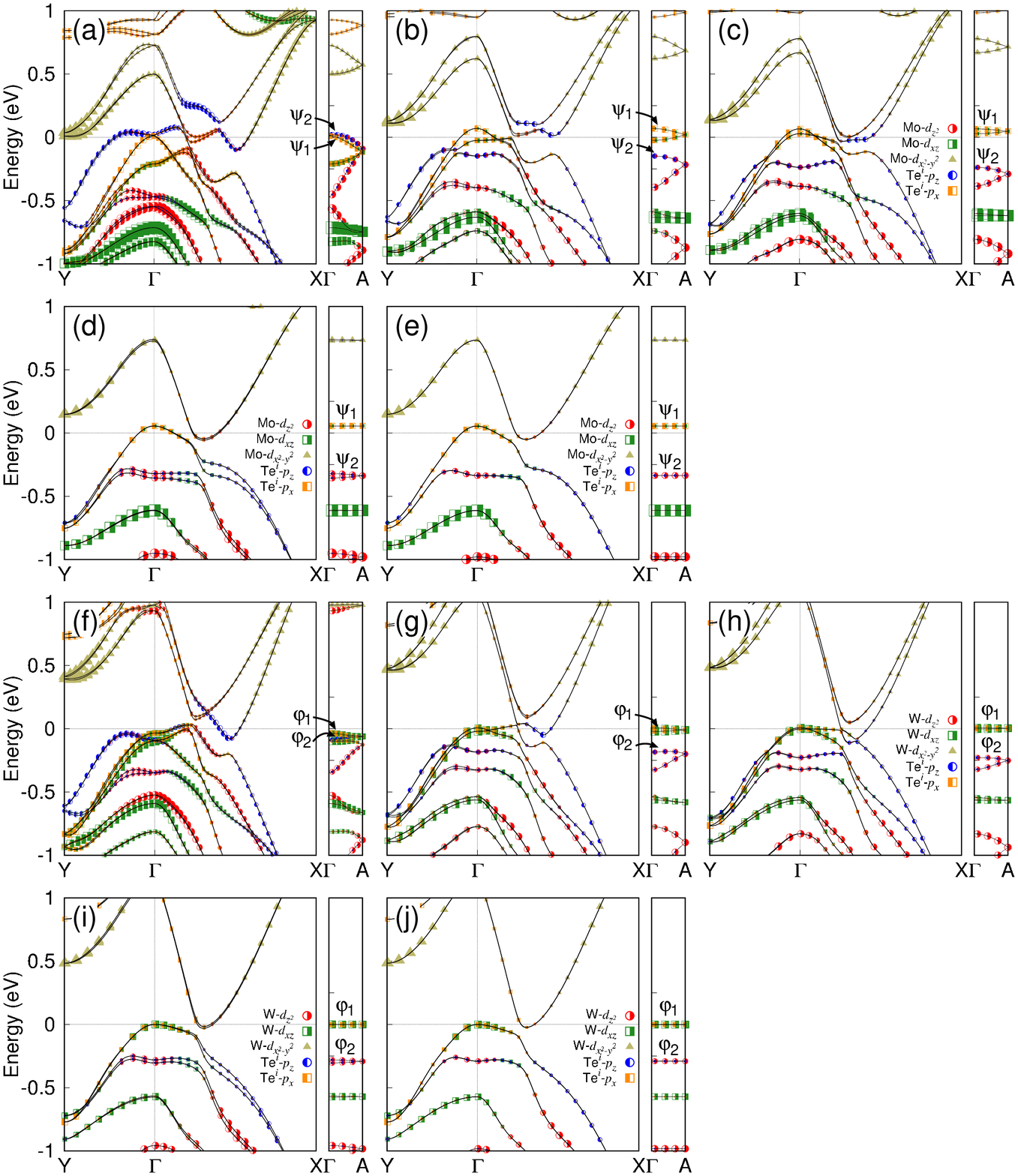} }
\setlength{\belowcaptionskip}{-30pt}
\caption{(Color online) Band structures of the (a)-(e) $\gamma$-MoTe$_2$ and (f)-(j) $\gamma$-WTe$_2$ with increasing interlayer distance $h$. From left to right panel, $\Delta$$h$ = $h$ - $h_0$ = (a),(f) 0.0 \AA, (b),(g) 0.5 \AA, (c),(h) 1.0 \AA, (d),(i) 3.0 \AA, and (e),(j) 8.0 \AA, respectively, where $h_0$ denotes the optimized interlayer distance obtained using the rev-vdW-DF2 method with SOC. The Fermi energy ($E_F$) is set to zero. The band structures are plotted along the path $Y$(0,$\frac{1}{2}$,0)$\rightarrow$$\Gamma$(0,0,0)$\rightarrow$$X$($\frac{1}{2}$,0,0) and $\Gamma$(0,0,0)$\rightarrow$$A$(0,0,$\frac{1}{2}$). In the monolayer limit of $\Delta h$ = 8.0 \AA [(e) and (j)], the topmost valence band state [denoted by $\psi_1$ and $\varphi_1$ in (e) and (j), respectively] is hybridized state of $p_x$ orbital of the lower Te atom [denoted by Te$^i$ in Fig. 1 of the main text] and $d_{z^2}$ orbital of transition metal atom. The next valence band state [$\psi_2$ and $\varphi_2$ in (e) and (j), respectively] is a hybridization of $p_z$ orbital of Te$^i$ and $d_{xz}$ orbital of transition metal atom. As decreasing interlayer distance from the monolayer limit to the bulk ($\Delta h$ = 0.0 \AA) limit, these states are further hybridized due to the interlayer interaction and split into bonding and anti-bonding states along the $\Gamma$-$A$ direction. It is noteworthy that the $p_x$ and $p_z$ orbital character in $\psi_1$ and $\psi_2$, respectively, are much stronger in MoTe$_2$ than those orbital character of $\varphi_1$ and $\varphi_2$ in WTe$_2$. This explains the smaller band width (along $\Gamma$-$A$) in WTe$_2$ compared with the width in MoTe$_2$. Most importantly, since the $p_z$ orbital character in the $\psi_2$ state, the splitting is much larger than that of the $\psi_1$. As a result, for MoTe$_2$, the anti-bonding $\psi_2$ state with $p_z$ orbital becomes the topmost valence band state as seen in (a). For WTe$_2$, however, owing to the reduced $p$ orbital contribution compared to that of MoTe$_2$, $\varphi_1$ and $\varphi_2$ are not reversed in the bulk limit as shown in (f).}
\end{figure}

\begin{figure}[h]
\centering{ \includegraphics[width=14.0cm]{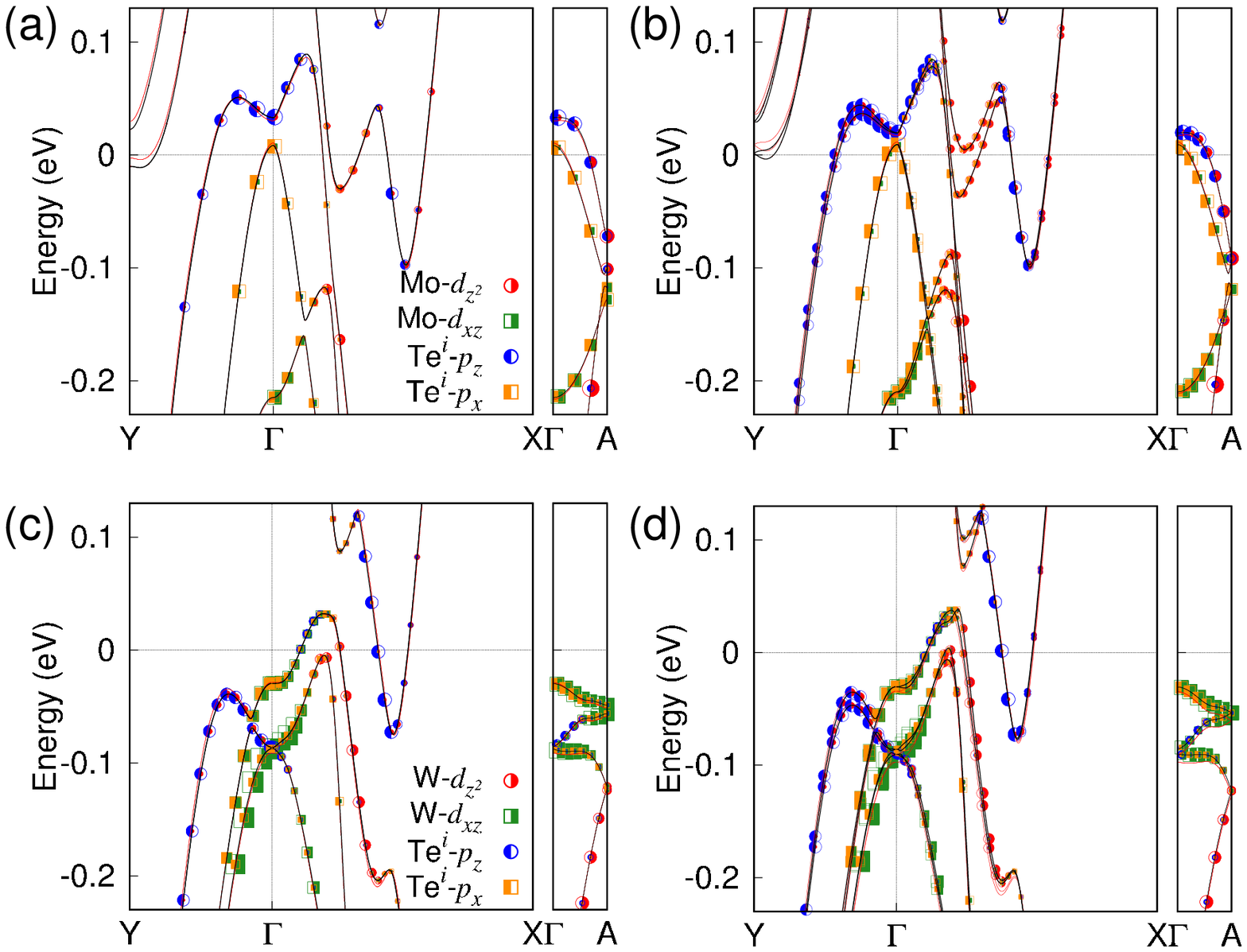} }
\caption{(Color online) Band structures of the (a) $\beta$-MoTe$_2$, (b) $\gamma$-MoTe$_2$, (c) $\beta$-WTe$_2$, and (d) $\gamma$-WTe$_2$ calculated with PBE exchange correlation functionals. Here, the geometries optimized by rev-vdW-DF2 method is taken. The Fermi energy ($E_F$) is set to zero. The band structures are plotted along the path $Y$(0,$\frac{1}{2}$,0)$\rightarrow$$\Gamma$(0,0,0)$\rightarrow$$X$($\frac{1}{2}$,0,0) and $\Gamma$(0,0,0)$\rightarrow$$A$(0,0,$\frac{1}{2}$). For the comparison, band structure of the rev-vdW-DF2 method is also plotted by red solid lines. There is no significant changes in the band structures especially in the valence states. }
\end{figure}

\begin{figure}[h]
\centering{ \includegraphics[width=16.0cm]{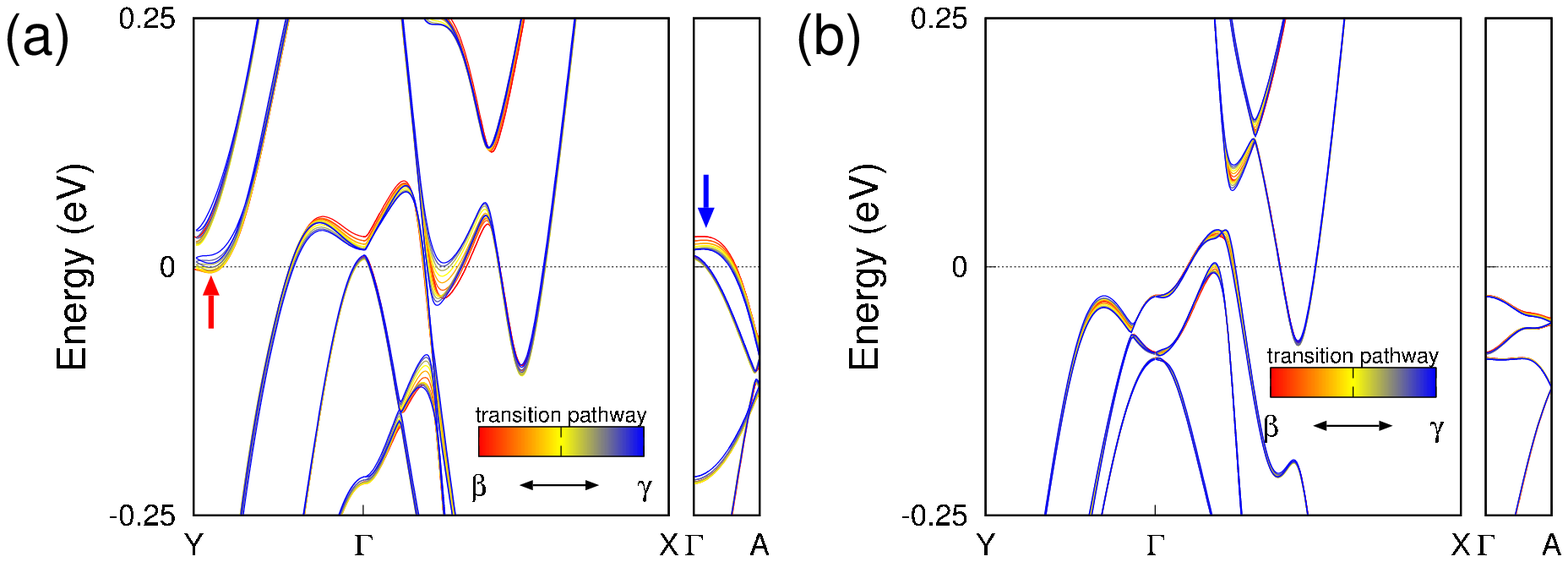} }
\caption{(Color online) Band structures evolution of the (a) MoTe$_2$ and (b) WTe$_2$ along the transition pathway from the $\beta$ to the $\gamma$ phase. The Fermi energy ($E_F$) is set to zero. The band structures are plotted along the path $Y$(0,$\frac{1}{2}$,0)$\rightarrow$$\Gamma$(0,0,0)$\rightarrow$$X$($\frac{1}{2}$,0,0) and $\Gamma$(0,0,0)$\rightarrow$$A$(0,0,$\frac{1}{2}$). For MoTe$_2$, there is a minute but considerable change in the band structures with the transition. That is, along the phase transition, the partially occupied valence bands related with the interlayer anti-bonding states [depicted with the blue arrow in (a)] shift down in energy while the intralayer bonding states [depicted with the red arrow in (a)] becomes to be unoccupied. So, there is a net charge transfer from the intralayer bonding states (mostly consists of $d_{x^2-y^2}$ orbital of Mo atoms, see Fig. S2) to interlayer anti-bonding states as increasing $d_1$ along the transition. This process can stabilize the anti-bonding state with increased bond distance. For the WTe$_2$, those bands are fully occupied and the band structures does not change significantly along the transition owing to the reduced interlayer interaction as discussed in Fig. S2.
}
\end{figure}

\begin{figure}[ht]
\centering{ \includegraphics[width=14.0cm]{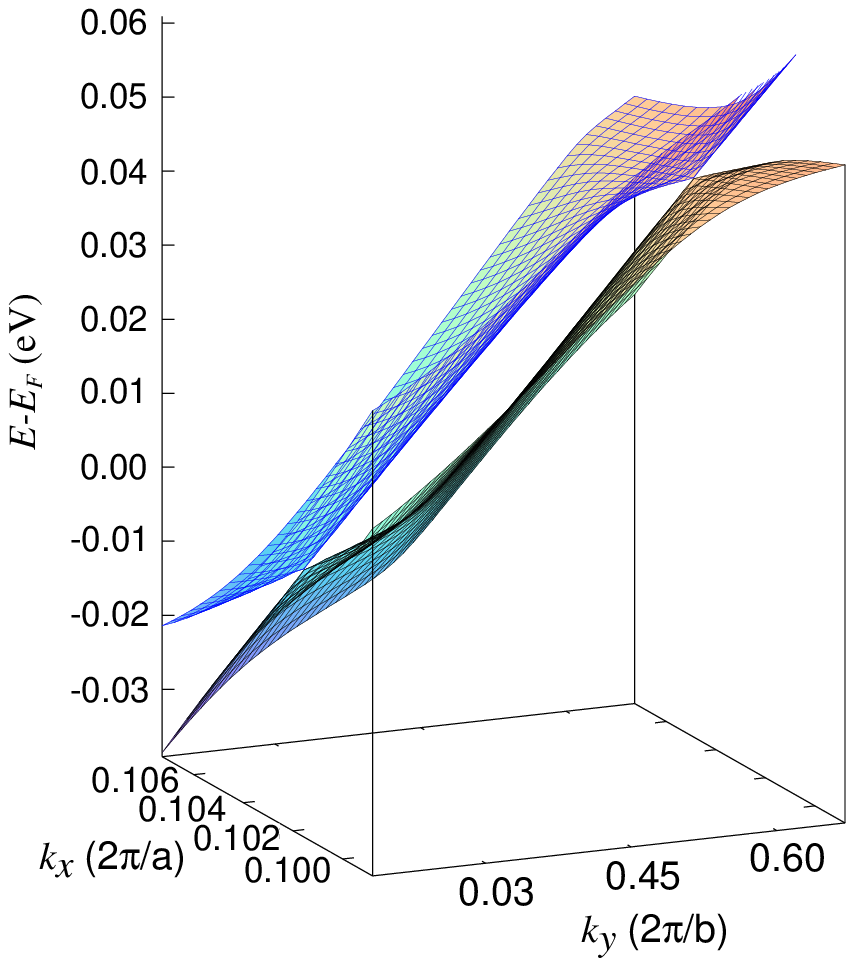} }
\caption{(Color online) 3D plot of the valence and conduction band edge states in the first quadrant with $k_z$ = 0 plane of the Brillouin zone, showing two Weyl points located at (0.030$\frac{2\pi}{b}$,0.102$\frac{2\pi}{a}$,0) and (0.053$\frac{2\pi}{b}$,0.101$\frac{2\pi}{a}$,0), respectively. Calculated locations of the Weyl points agree with those in a previous report very well~\cite{Sun2015}. The Fermi energy ($E_F$) is set to zero.}
\end{figure}

\begin{figure}[h]
\centering{ \includegraphics[width=12.0cm]{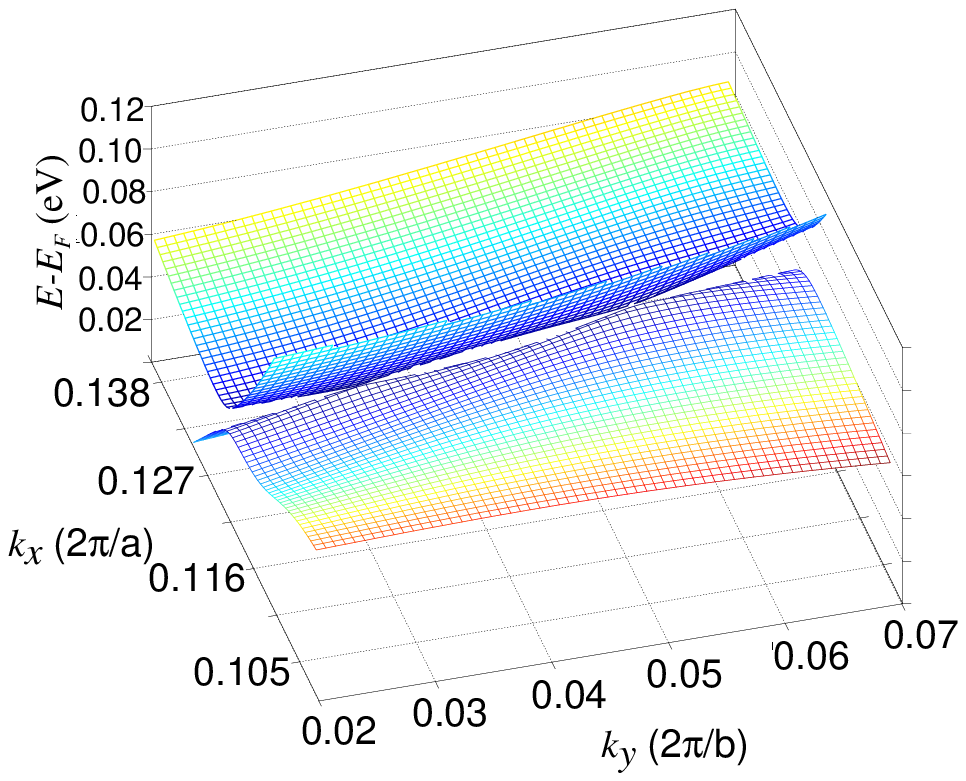} }
\caption{(Color online) 3D plot of the valence and conduction band edge states of strained ($-$1.5$\%$ along $\bf a$ and $\bf c$) $\gamma$-WTe$_2$ structure calculated by using the rev-vdW-DF2 method. The first quadrant of Brillouin zone of $k_z$ = 0 plane is plotted to show two Weyl points located at (0.031$\frac{2\pi}{b}$,0.122$\frac{2\pi}{a}$,0) and (0.055$\frac{2\pi}{b}$,0.120$\frac{2\pi}{a}$,0), respectively. We also note there are eight Weyl points in the full Brillouin zone with $k_z$ = 0 plane. The Fermi energy ($E_F$) is set to zero.}
\end{figure}

\newpage

\begin{table}[ht]
\caption{Calculated lattice parameters of the $\beta$-MoTe$_2$ and $\gamma$-MoTe$_2$, obtained by using the various exchange correlation functionals with spin-orbit coupling (SOC). Experiment values for the lattice parameters are also shown for comparison. The corresponding values without SOC are also given in parentheses.}
\begin{ruledtabular}
\begin{tabular}{lccccccc}
            & a(\AA) & b(\AA) & c(\AA) & $\theta(^{\circ})$ \\  
\cline{2-5}
 $\beta$-MoTe$_2$\\
 \cline{1-5}
 LDA          &3.440 (3.436)& 6.292 (6.300) & 13.361 (13.317) &93.580 (93.593) \\
 PBE          &3.469 (3.456)& 6.366 (6.369) & 15.636 (15.642) &91.681 (91.965) \\
 PBE+D2       &3.496 (3.492)& 6.388 (6.394) & 13.742 (13.712) &93.472 (93.478) \\
 PBE+D3       &3.470 (3.467)& 6.359 (6.366) & 13.599 (13.541) &93.501 (93.550) \\
 vdW-DF       &3.580 (3.562)& 6.442 (6.444) & 15.023 (15.116) &90.962 (91.446) \\
 vdW-DF2      &3.667 (3.655)& 6.543 (6.540) & 14.887 (14.934) &91.650 (91.640) \\
 optB88-vdW   &3.531 (3.525)& 6.360 (6.361) & 14.158 (14.146) &92.719 (92.312) \\
 rev-vdW-DF2  &3.498 (3.495)& 6.331 (6.337) & 13.868 (13.819) &93.369 (93.423) \\
 Experiment\footnote{Reference~\cite{Tamai2016}, X-ray diffraction study (250 K) \label{a}} & 3.479 & 6.332  &13.832   &93.830 \\ 
 Experiment\footnote{Reference~\cite{Keum2015}, X-ray diffraction study \label{a}} & 3.475 & 6.3274 &13.8100  &93.887 \\ 
\\ 
 $\gamma$-MoTe$_2$\\
 \cline{1-4}
 LDA          & 3.443 (3.440) & 6.289 (6.296)& 13.341 (13.302)\\
 PBE          & 3.471 (3.457) & 6.364 (6.368)& 15.410 (15.569)\\
 PBE+D2       & 3.501 (3.497) & 6.378 (6.384)& 13.710 (13.680)\\
 PBE+D3       & 3.472 (3.470) & 6.356 (6.362)& 13.583 (13.555)\\
 vdW-DF       & 3.580 (3.563) & 6.442 (6.443)& 15.022 (15.115)\\
 vdW-DF2      & 3.668 (3.655) & 6.543 (6.539)& 14.879 (14.920)\\
 optB88-vdW   & 3.532 (3.527) & 6.356 (6.359)& 14.151 (14.131)\\
 rev-vdW-DF2  & 3.497 (3.495) & 6.327 (6.333)& 13.878 (13.828)\\
 Experiment\footnote{Reference~\cite{Tamai2016}, X-ray diffraction study (100 K) \label{b}} & 3.468  & 6.310 & 13.861  \\ 
 Experiment\footnote{Reference~\cite{Wang2016}, X-ray diffraction study (100 K) \label{b}} & 3.4582 & 6.3043& 13.859  \\ 
\end{tabular}
\end{ruledtabular}
\end{table}

\begin{table}[ht]
\caption{Calculated lattice parameters of the $\beta$-WTe$_2$ and $\gamma$-WTe$_2$, obtained by using the various exchange correlation functionals with SOC. Experiment values for the lattice parameters are also shown for comparison. The corresponding values without SOC are also given in parentheses.}
\begin{ruledtabular}
\begin{tabular}{lccccccc}
            & a(\AA) & b(\AA) & c(\AA) & $\theta(^{\circ})$ \\  
\cline{2-5}
 $\beta$-WTe$_2$\\
 \cline{1-5}
 LDA         &   3.453 (3.445)&  6.210 (6.214)&  13.810 (13.809) & 92.217 (92.287) \\ 
 PBE         &   3.507 (3.494)&  6.312 (6.314)&  15.647 (15.800) & 91.426 (91.360) \\
 PBE+D2      &   3.543 (3.533)&  6.227 (6.231)&  13.764 (13.772) & 91.094 (91.128) \\
 PBE+D3      &   3.479 (3.464)&  6.282 (6.283)&  14.029 (14.104) & 92.004 (91.870) \\
 vdW-DF      &   3.597 (3.584)&  6.392 (6.392)&  15.188 (15.256) & 90.828 (91.138) \\
 vdW-DF2     &   3.681 (3.665)&  6.484 (6.482)&  15.042 (15.101) & 91.267 (90.663) \\
 optB88-vdW  &   3.545 (3.533)&  6.310 (6.310)&  14.363 (14.406) & 90.601 (90.394) \\
 rev-vdW-DF2 &   3.508 (3.496)&  6.278 (6.279)&  14.173 (14.221) & 92.052 (92.086)\\
 \\
 $\gamma$-WTe$_2$\\
 \cline{1-4}
 LDA         &  3.454 (3.445)& 6.208 (6.211)& 13.818 (13.832) \\
 PBE         &  3.507 (3.495)& 6.311 (6.314)& 15.698 (15.672) \\
 PBE+D2      &  3.544 (3.534)& 6.225 (6.230)& 13.761 (13.767) \\
 PBE+D3      &  3.479 (3.463)& 6.279 (6.282)& 14.042 (14.143) \\
 vdW-DF      &  3.597 (3.585)& 6.392 (6.391)& 15.162 (15.232) \\
 vdW-DF2     &  3.681 (3.665)& 6.484 (6.482)& 15.023 (15.103) \\
 optB88-vdW  &  3.545 (3.533)& 6.309 (6.310)& 14.363 (14.419) \\
 rev-vdW-DF2 &  3.509 (3.498)& 6.276 (6.278)& 14.171 (14.217)  \\
Experiment\footnote{Reference~\cite{Brown1966}, X-ray diffraction study \label{brown}}& 3.496 & 6.282 & 14.07 &         \\
Experiment\footnote{Reference~\cite{Mar1992}, X-ray diffraction study (113 K) \label{mar}}& 3.477 & 6.249 & 14.018 &         \\
Experiment\footnote{Reference~\cite{Pan2015}, X-ray diffraction study (296 K) \label{pan}}& 3.486 & 6.265 & 14.038 &         \\
\end{tabular}
\end{ruledtabular}
\end{table}

\begin{table}[ht]
\caption{The calculated interatomic distances (in \AA) of MoTe$_{2}$ and WTe$_{2}$ using the
rev-vdW-DF2 method including SOC effect, in comparison with the experimental data. 
The corresponding values obtained without SOC are also given in parentheses.
The labeling of each atoms is shown below.}
\begin{ruledtabular}
\begin{tabular}{cccccccc}
	& \multicolumn{7}{c}{\includegraphics[height=7cm]{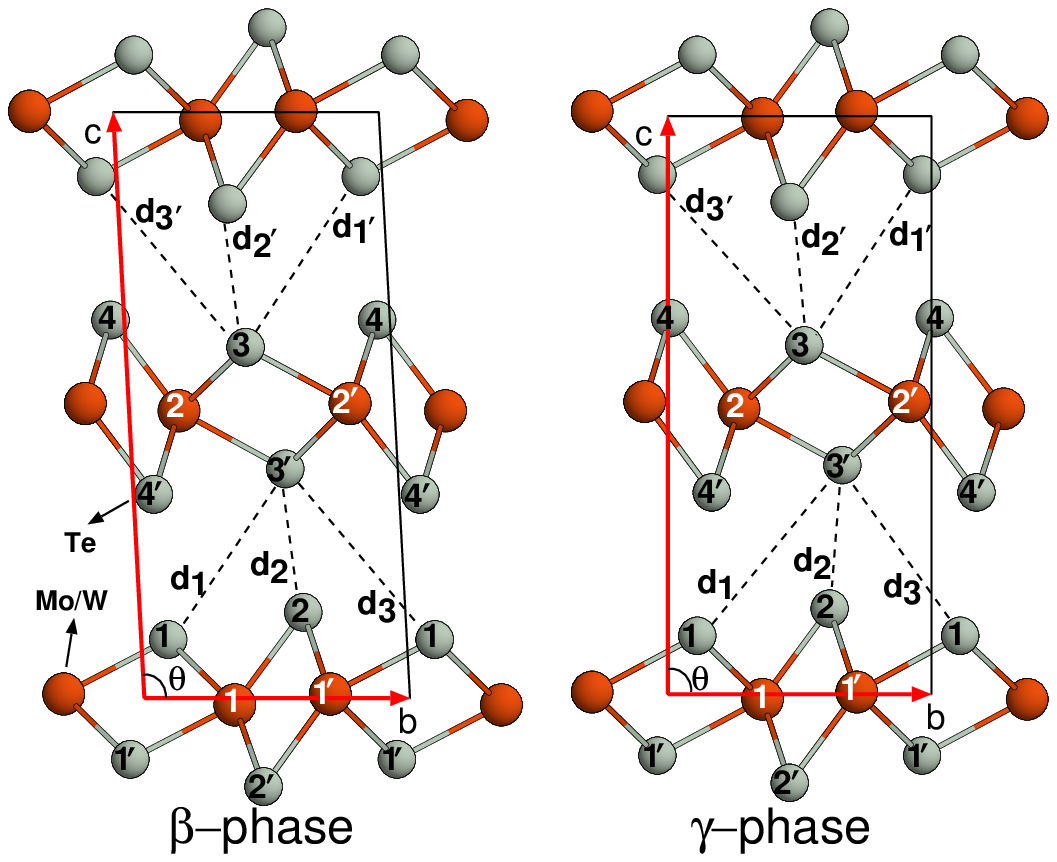}}\\
        & \multicolumn{4}{c}{MoTe$_{2}$}  & \multicolumn{3}{c}{WTe$_{2}$} \\
\cline{2-5}
\cline{6-8}
 & \multicolumn{2}{c}{This work } & \multicolumn{2}{c}{Experiment}  &  \multicolumn{2}{c}{This work} & \multicolumn{1}{c}{Experiment} \\
\cline{2-3}
\cline{4-5}
\cline{6-7}
\cline{8-8}
      &     $\beta$-phase        &  $\gamma$-phase        & $\beta$-phase~\footnote{Reference~\cite{Brown1966}}& $\gamma$-phase~\footnote{Reference~\cite{Qi2016}} & $\beta$-phase & $\gamma$-phase & $\gamma$-phase~\footnote{Reference~\cite{Mar1992}} \\
M$_{1}$$-$M$_{1'}$   &   2.874 (2.875)& 2.871 (2.872) & 2.890  & 2.898  & 2.838 (2.835) &  2.837 (2.834)& 2.849  \\
M$_{2}$$-$M$_{2'}$   &   2.872 (2.873)& 2.871 (2.872) & 2.901  & 2.898  & 2.837 (2.835) &  2.837 (2.834)& 2.849  \\
M$_{1}$$-$Te$_{1}$   &   2.811 (2.809)& 2.813 (2.808) & 2.786  & 2.806  & 2.824 (2.825) &  2.825 (2.825)& 2.800  \\
M$_{1}$$-$Te$_{2}$   &   2.712 (2.711)& 2.719 (2.710) & 2.701  & 2.715  & 2.732 (2.732) &  2.731 (2.732)& 2.712  \\
M$_{1'}$$-$Te$_{2}$  &   2.711 (2.710)& 2.709 (2.711) & 2.706  & 2.702  & 2.725 (2.722) &  2.726 (2.723)& 2.698  \\
M$_{1'}$$-$Te$_{1}$  &   2.826 (2.825)& 2.825 (2.824) & 2.818  & 2.814  & 2.835 (2.834) &  2.834 (2.833)& 2.803  \\
M$_{2}$$-$Te$_{4'}$  &   2.710 (2.709)& 2.713 (2.708) & 2.691  & 2.705  & 2.725 (2.721) &  2.724 (2.721)& 2.699  \\
M$_{2}$$-$Te$_{3'}$  &   2.825 (2.824)& 2.825 (2.824) & 2.801  & 2.817  & 2.834 (2.833) &  2.834 (2.833)& 2.802  \\
M$_{2'}$$-$Te$_{3'}$ &   2.812 (2.811)& 2.810 (2.811) & 2.789  & 2.806  & 2.825 (2.826) &  2.825 (2.826)& 2.798  \\
M$_{2'}$$-$Te$_{4'}$ &   2.717 (2.715)& 2.712 (2.717) & 2.711  & 2.705  & 2.735 (2.734) &  2.735 (2.735)& 2.705  \\
$d_{1}$              &   4.889 (4.874)& 5.360 (5.336) & 4.846  & 5.435  & 5.068 (5.077) &  5.339 (5.349)& 5.397  \\
$d_{2}$              &   3.854 (3.837)& 3.860 (3.842) & 3.862  & 3.869  & 3.945 (3.958) &  3.943 (3.955)& 3.911  \\
$d_{3}$              &   5.402 (5.396)& 4.947 (4.948) & 5.443  & 4.845  & 5.379 (5.392) &  5.111 (5.122)& 4.937  \\
$d_{1'}$             &   4.889 (4.874)& 4.947 (4.948) & 4.846  & 4.845  & 5.068 (5.077) &  5.111 (5.122)& 4.937  \\
$d_{2'}$             &   3.854 (3.837)& 3.860 (3.842) & 3.862  & 3.869  & 3.945 (3.958) &  3.943 (3.955)& 3.911  \\
$d_{3'}$             &   5.402 (5.396)& 5.360 (5.336) & 5.443  & 5.435  & 5.379 (5.393) &  5.339 (5.349)& 5.397  \\
\end{tabular}
\end{ruledtabular}
\end{table}

\end{document}